\documentclass[final,journal,10pt]{IEEEtran}
\usepackage{amsmath}
\usepackage{amsthm}
\usepackage{cite}
\usepackage{graphicx}
\usepackage{subfigure}
\usepackage{epstopdf}
\usepackage{amsfonts,amsmath,amssymb}
\usepackage{graphicx}
\usepackage{bm}
\usepackage{color}
\usepackage{caption}
\usepackage{multirow}
\usepackage{makecell}
\usepackage{enumitem}
\usepackage{longtable}

\usepackage{setspace}

\ifCLASSINFOpdf
\else
\fi
\begin{document}
%
\title{\huge{An Integrated Optimization-Learning Framework for Online Combinatorial Computation Offloading in MEC Networks}}
%
%
%

\author{Xian~Li,
        Liang~Huang,
        Hui~Wang,
        Suzhi~Bi,
        and~Ying-Jun Angela~Zhang}
\maketitle

\begin{abstract}
Mobile edge computing (MEC) is a promising paradigm to accommodate the increasingly prosperous delay-sensitive and computation-intensive applications in 5G systems. To achieve optimum computation performance in a dynamic MEC environment, mobile devices often need to make online decisions on whether to offload the computation tasks to nearby edge terminals under the uncertainty  of future system information (e.g., random wireless channel gain and task arrivals). The design of an efficient online offloading algorithm is challenging. On one hand, the fast-varying edge environment requires frequently solving a hard combinatorial optimization problem where the integer offloading decision and continuous resource allocation variables are strongly coupled. On the other hand, the uncertainty of future system parameters makes it hard for the online decisions to satisfy long-term system constraints. To address these challenges, this article overviews the existing methods and introduces a novel framework that efficiently integrates model-based optimization and model-free learning techniques. Besides, we suggest some promising future research directions of online computation offloading control in MEC networks.
\end{abstract}

\section{Introduction}
\begin{figure}
	\centering
	\includegraphics[scale=0.31]{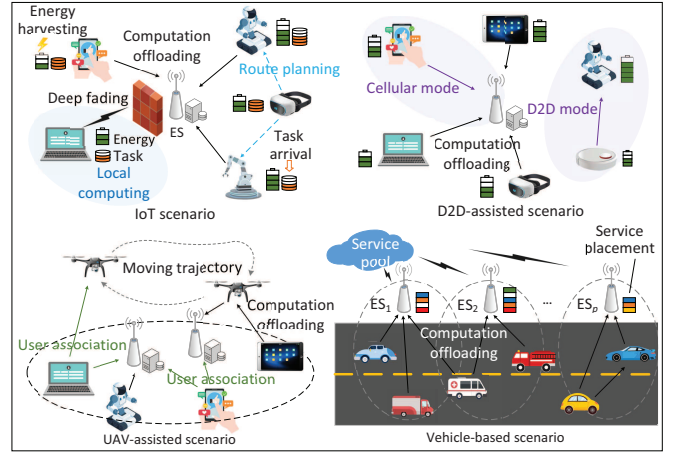}
	\captionsetup{font=footnotesize}
	\caption{Example computation offloading scenarios of MEC system.}
	\label{fig1}
\end{figure}

To meet the stringent latency requirements of thriving wireless intelligent applications (e.g., online 3D gaming and autonomous driving), mobile edge computing (MEC) \cite{MaoFourthquarter2017} has emerged as a new computation paradigm to provide ubiquitous computation services. Via pushing computation resources toward network edges, MEC enables flexible and rapid deployment of new services that require intensive computation and low latency. In MEC systems, mobile devices (MDs) with hardware limitations (e.g., limited local computational power, storage and energy) can offload their computation-intensive tasks to nearby resourceful edge servers. As shown in Fig. \ref{fig1}, the potential application of MEC technology ranges from typical IoT (internet of things) networks to newly-emerging UAV (unmanned aerial vehicle) communication systems.

Rather than persistently offloading tasks for edge processing, offloading computation tasks in an opportunistic manner offers substantial performance gain. For instance, MDs may prefer local computing to offloading to a server when the communication channel to the edge server (ES) is weak or the ES is heavily loaded. Due to the sharing of limited edge communication and computation resources among MDs, the binary offloading decisions of different MDs or tasks (e.g., offloading a computation task or computing locally) are strongly coupled and shall be jointly optimized. As a result, the optimal offloading solution needs to be selected from a discrete action set that grows exponentially with the problem size (e.g., number of MDs or tasks). Besides, the continuous resource allocation variables (e.g., CPU frequency, transmit power and bandwidth) need to be jointly optimized with the integer offloading decisions. This renders the computation offloading problem (COP) a mixed-integer nonlinear programming (MINLP) that is generally hard to be solved efficiently.\

In practice, the volatile MEC environment urges MDs to perform online offloading decisions in real time under the uncertainty of future system parameters (e.g., random channel conditions and task arrivals). The corresponding online offloading algorithm design faces two challenges. First, the fast-varying edge environment requires frequently re-solving the hard COP when the system parameters vary significantly. When the problem size is large, it is computationally prohibitive to obtain the optimal offloading solution in real-time implementation. Second, with uncertain future system information, it is hard for the online offloading decisions to satisfy long-term system constraints, e.g., average power consumption constraint or achieving task data queue stability.

Existing online offloading algorithms are mainly categorized into model-based optimization and model-free machine learning algorithms. In particular, the former formulates and solves the COP based on explicit mathematical models that describe the physical system. A major advantage of model-based optimization is the theoretical guarantee of convergence toward an optimal offloading solution. Due to the combinatorial nature of COP, optimization-based methods often require a large number of iterations to achieve a satisfying solution. Besides, they fail to exploit the experience learned from historical data, and hence need to recalculate the COP from scratch once system parameters change. As a result, it is fundamentally difficult for optimization-based methods to adapt to fast-varying MEC environments.\

Recently, machine learning based methods, especially deep reinforcement learning (DRL), have emerged as promising alternatives to mathematical optimization. Based on the past experience, DRL takes advantage of deep neural networks (DNNs) to construct a direct mapping from the input system parameters to the output offloading and resource allocation decisions without explicitly solving the mathematical optimization of COP. Its low execution latency is particularly appealing to online decision making. A key drawback of model-free DRL is the slow convergence when the dimension of the action space is large. The mixed integer-continuous variables in COP yield a very large, or even infinite, action space, yielding the curse-of-dimensionality. {\color{black}{Besides, the actions produced by the native action-reward structure of DRL may not satisfy the constraints of COP, such as per-slot SINR (signal-to-interference-plus-noise ratio) constraint and long-term queue stability constraint.}} \

In this article, we introduce a novel framework that skillfully integrates model-based optimization and model-free DRL to overcome their respective drawbacks mentioned above. We adopt an actor-critic structure (as illustrated in Fig. \ref{fig3} in section \ref{sec3}) to optimize the offloading and resource allocation decisions. Specifically, we apply a DNN-based actor module to generate integer offloading decisions and resort to conventional optimization in the critic module to obtain the continuous resource allocation solution given the generated offloading decisions. The two modules operate alternately: the critic module evaluates the integer offloading decision produced by the actor module, and the actor module uses the evaluations to update its mapping policy for generating the offloading decisions. Such a training process exploits the past experience to repeatedly update the DNN model parameters (i.e., the policy of the DNN in the actor module) until converging to the optimal mapping policy. Subsequently, we can directly map any new input system parameters to the optimal output offloading decisions without numerical optimization, taking negligible computation time against model-based optimization methods. This essentially facilitates real-time online decision making in a fast-varying MEC environment. On the other hand, compared to conventional DRL that treats both the integer offloading and continuous resource allocation decisions as the action, the proposed approach significantly reduces the action space of the actor module by considering only the binary offloading decisions as the action, leaving resource allocation to the numerical optimization in the critic module. This greatly simplifies the learning task to a classical classification problem. More importantly, the optimization-based critic module provides precise evaluation of the integer offloading decisions generated by the actor module. This greatly improves the convergence of the training process in comparison with conventional DRL, whose convergence is often jeopardized by the imprecise evaluation on actions before the critic network is sufficiently trained. Moreover, the optimization-based critic guarantees the feasibility of the resource allocation solutions. Last but not least, the proposed framework can flexibly incorporate various off-the-shelf learning and optimization methods according to different MEC application requirements. For instance, by introducing the well-established Lyapunov optimization technique, we are able to handle long-term performance constraints under random system parameters, as detailed in section \ref{sec4}. Overall, the proposed integrated optimization-learning framework employs the benefits of both sides. By learning from the past experience, it is able to generate optimal offloading solutions very quickly, greatly enhancing the practicality in fast-varying MEC environments. Meanwhile, model-based optimization in the critic module guarantees fast and robust convergence even when the action space is large.


The article is organized as follows. In section \ref{sec2}, we introduce the online computation offloading problem in MEC networks and review some conventional methods. In section \ref{sec3}, we introduce the novel integrated optimization-learning framework for online offloading control and discuss in section \ref{sec4} its extension to stochastic cases. At last, we suggest some future research directions and conclude the paper.

\section{Online Computation Offloading Problem and Conventional Solutions}\label{sec2}
We consider an MEC network consisting of one or multiple ESs and MDs. Depending on the application scenario, the MDs can be powered by grid power, batteries, or harvested energy. Suppose that the time is divided into consecutive slots, within each the system parameters (such as wireless channel condition) remain unchanged. In each time slot, the system operator makes computation offloading decision for each task based on the current system parameters. We adopt a binary computation offloading policy \cite{MaoFourthquarter2017}, namely, the tasks are either computed locally or at the ES. In general, an online computation offloading problem contains the following three key elements.\

\begin{itemize}[leftmargin=*]
	\item \textbf{Problem Parameters}: The problem parameters include exogenous and endogenous parameters. In time slot $t$, the exogenous parameters, denoted as $\bm{o}^t$, capture the realizations of exogenous environment random variables, including the wireless channel gain, task data arrivals, and harvested renewable energy, etc. On the other hand, the endogenous parameters, denoted as $\bm{I}^t$, depict the time-varying system state of the ESs and the MDs, such as the battery level and data queue length, as well as the static hardware settings of the MEC system like the maximum CPU frequency and transmit power. For simplicity, we denote the combination of exogenous and endogenous parameters as $\bm{s}^t=\{\bm{o}^t,\bm{I}^t\}$.
		
	\item \textbf{Action}: The action in time slot $t$, denoted as $\bm{a}^t$, includes computation offloading and resource allocation decisions. In particular, $\bm{a}^t$ consists of both integer and continuous variables, denoted as $\bm{a}^t=\{\bm{x}^t, \bm{y}^t\}$. The integer variables $\bm{x}^t$ correspond to the binary offloading decisions and the continuous variables $\bm{y}^t$ correspond to the resource allocation decisions such as the transmit power and bandwidth occupied by MDs. {\color{black}{These actions are constrained by limited system resource and users’ performance requirements in each time slot. Besides, the sequential actions $\bm{a}^t$, for $t=0,1,\cdots$, may be subject to long-term constraints such as average power consumption and task data queue stability of the ESs and MDs.}}	
	
	\item \textbf{System Utility}: The system utility in the $t$th time slot, denoted as $r^t$, is a real-valued function $r^t=f(\bm{s}^t,\bm{a}^t)$ that measures the per-slot performance of the adopted action $\bm{a}^t$ under parameters $\bm{s}^t$. For example, in a computation-intensive application, the utility can be the data processing rate of the system. For energy-aware applications, it is the energy consumption of energy-scarce devices.
\end{itemize}

In a fast-varying edge environment, it is very challenging to obtain the optimal actions $\bm{a}^t$ in real time due to the combinatorial nature of the problem. In the following, we provide a brief review of existing offloading algorithms in MEC systems.\

\subsection{Model-Based Optimization}
Optimization-based methods apply integer or convex optimization techniques to solve the COP based on explicit mathematical formulations (i.e., model-based). Some representative approaches include:
\begin{itemize}[leftmargin=*]
	\item \textbf{Relaxation-based approaches}: The main idea is to first relax the integer variables to continuous ones and reformulate the problem into an ``easier'' continuous optimization problem. Then, a feasible but often suboptimal integer solution is recovered from the solution of the relaxed problem. The relaxation can be performed either directly on binary offloading decision variables through linear relaxation (LR) \cite{Wang2019} and semidefinite relaxation (SDR) \cite{Dinh2017}, or indirectly on some key metrics related to the binary decisions, e.g., the portion of energy consumed by computation offloading \cite{Bi2018a}. Despite the low computing complexity, the solution quality of relaxation-based methods is not guaranteed. Besides, some MINLP problems after relaxation are still hard non-convex optimization problems due to the strong coupling between the integer and continuous variables.
		
	\item \textbf{Local-search-based approaches}: Local-search-based approaches start from an initial solution and iteratively find a better solution in the vicinity of the current solution until no further improvement can be made. For example, \cite{Bi2018a} applies the coordinate descent (CD) method to swap the computation mode (e.g., local computing or edge computing) of only one best MD in each iteration until a local optimum is reached. To avoid trapping into a local optimum, \cite{Yan2020} uses Gibbs sampling to update the binary variables according to a specific probability distribution function.
	
	\item \textbf{Metaheuristic approaches}: Instead of enumerating all the possible combinations, metaheuristics search over a subset of feasible solutions in a structured manner, e.g., using bio-inspired method. For example, \cite{Guo2018} proposed a hierarchical algorithm based on genetic algorithm (GA) and particle swarm optimization (PSO) to optimize the combinatorial offloading decisions in a large action space.
	
	\item \textbf{Decomposition approaches}: Decomposition methods effectively reduce the computational complexity of mixed-integer optimization by decomposing a large-size problem into smaller parallel sub-problems. Blending the benefits of dual decomposition and augmented Lagrangian methods, alternating direction method of multipliers (ADMM) decomposes the hard combinatorial COP into parallel subproblems with only one integer variable \cite{XianIWCL2020}. The optimal offloading decision of each subproblem is obtained by simply enumerating two individual offloading options, i.e., offloading and local computation, for each user.
\end{itemize}

Due to the NP-hardness of COP, model-based optimization methods require a large number of iterations to reach a satisfying solution. Besides, they fail to make use of the valuable historical data, and thus need to completely recalculate the COP once the system parameters vary. As a result, the high complexity prohibits their real-time implementation in a fast-varying environment. \

\subsection{Model-Free DRL-based Methods}
Instead of explicitly solving each new instance of COP numerically, DRL-based methods build on the past experience to construct a direct mapping from the problem parameters to the optimal control action. By model free, we mean DRL is data driven instead of relying on an explicit (and sometimes imprecise) mathematical model of the system. Once the deep learning network is fully trained by the past experience, DRL can output the control actions very quickly in response to a new set of input system parameters. Existing DRL methods are mainly categorized into value-based, policy-based, and hybrid approaches, as detailed below.
\begin{itemize}[leftmargin=*]
	\item \textbf{Value-based DRL}: Value-based DRL methods approximate the state-action value function (generally referred to as Q-value) using a deep neural network (DNN) and select the most-rewarding action in each iteration based on the Q-value estimation of all feasible actions. Some recent application of value-based methods on online offloading control include deep Q-learning network (DQN) \cite{Min2019} and double DQN \cite{Chen2019}.
		
	\item \textbf{Policy-based DRL}: Policy-based DRL methods employ a DNN (referred to as the actor network) to construct the optimal offloading policy by iteratively tuning the DNN parameters using the policy gradient technique. Meanwhile, the quality of the policy is evaluated by a dedicated critic network using a separate DNN or Monte Carlo based methods. Commonly-used policy-based DRL methods in MEC systems include the actor-critic DRL \cite{Wei2019} and the deep deterministic policy gradient (DDPG) \cite{Dai2020}.
	
	\item \textbf{Hybrid DRL}: Hybrid DRL methods combine the merits of both value-based and policy-based DRL. For example, \cite{Zhang2020} adopts a DDPG-based actor module to obtain continuous resource allocation, followed by a DQN-based critic module to evaluate the resource allocation decision and select the best integer offloading decision.
\end{itemize}

\begin{figure*}
	\centering
	\includegraphics[scale=0.52]{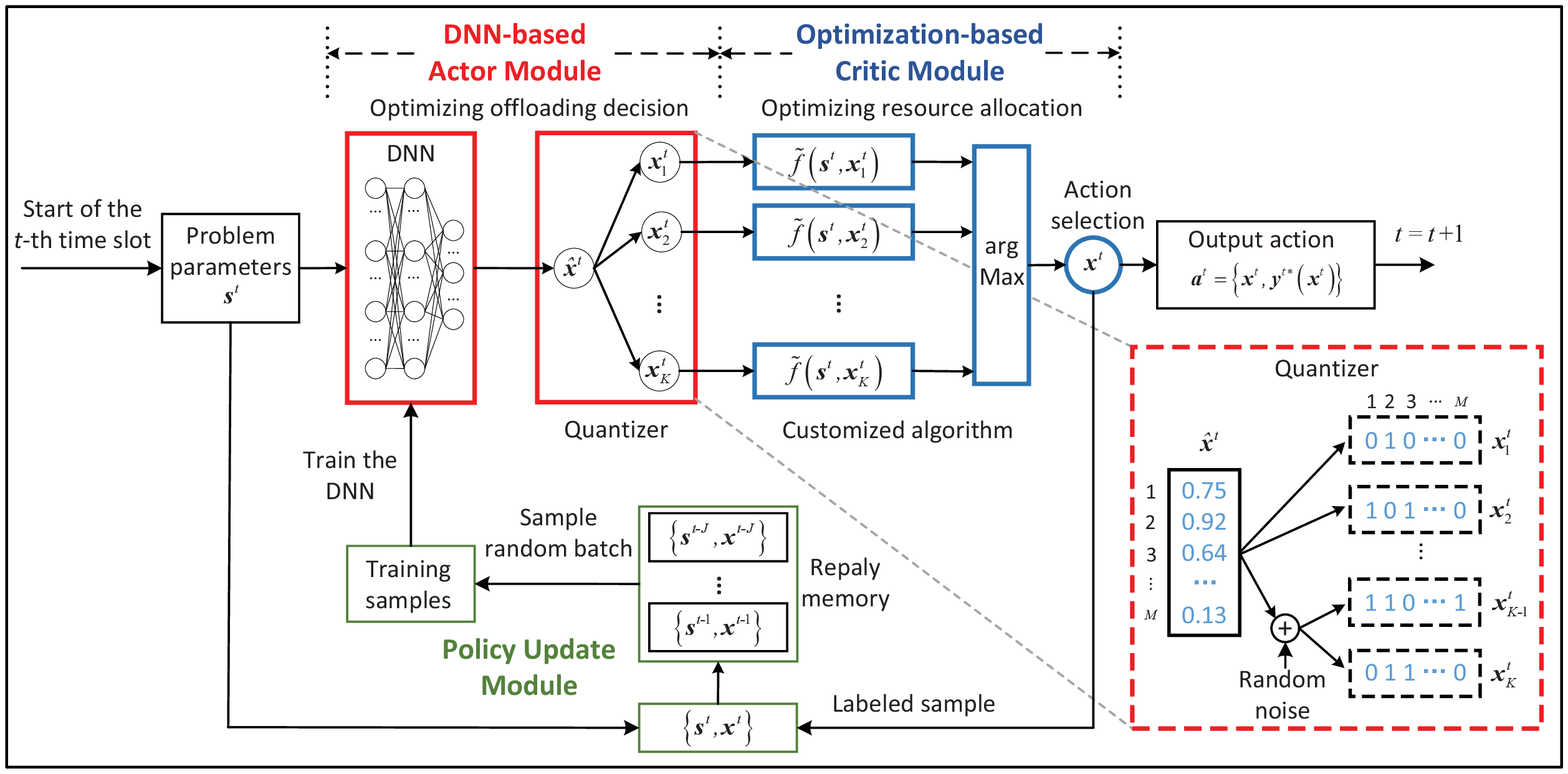}
	\captionsetup{font=footnotesize}
	\caption{The schematics of the DROO method.}
	\label{fig3}
\end{figure*}

To solve COP, value-based methods are computationally inefficient due to the large and hybrid integer-continuous action space. Also, it is frequently reported that policy-based methods suffer slow convergence or even divergence, especially when the critic module fails to produce a precise and stable value function approximation. Besides, model-free DRL cannot guarantee that the output actions satisfy both the per-slot and long-term system constraints.

To address the fundamental drawbacks of the existing optimization- and DRL-based methods, we introduce a novel integrated optimization-learning framework, referred to as DROO (deep reinforcement learning-based online offloading) \cite{Huang2020}, in the following section.\

\section{An Integrated Optimization and Learning Approach}\label{sec3}
In this section, we introduce the basic structure of DROO that solves a per-slot deterministic COP. For each slot $t$, we maximize the per-slot utility $f(\bm{s}_t,\bm{a}_t)$ given the realization of $\bm{s}_t$. In the next section, we will extend DROO to solve a general stochastic COP problem with long-term objective and constraints.\

As shown in Fig. \ref{fig3}, DROO solves the COP via three interactive modules, namely actor module, critic module, and policy update module. Specifically, the actor module adopts model-free learning, while the critic module adopts model-based optimization. Below are the details of the three modules.

\begin{itemize}[leftmargin=*]
	\item \textbf{Learning-based Actor Module}: The actor module consists of a DNN and an offloading action quantizer. It is different from the actor modules of conventional DRLs in the following two senses. First, instead of outputting an action $\bm{a}^t=\{\bm{x}^t, \bm{y}^t\}$, the proposed actor DNN outputs a relaxed offloading decision $\bm{\hat{x}}^t$ based on the input system parameters $\bm{s}^t$. This greatly reduces the action space of the DNN network and thus expediting the training process. Second, unlike the actor network of conventional policy-based DRL methods that generate only a single action, DROO quantizes $\bm{\hat{x}}^t$ into $K> 1$ candidate binary offloading decisions $\bm x_j^t$ ($j=1, \cdots, K$), from which one best decision will be selected later. The quantization procedure unifies the exploration and exploitation during the learning progress. In particular, setting a larger $K$ leads to more exploration than exploitation, and vice versa. As a rule of thumb, a larger $K$ is to be used when device number is large. Intuitively, the $K$ $\bm x_j^t$'s should be sufficiently close to $\bm{\hat{x}}^t$ to make good use of the DNN's output, and meanwhile sufficiently separated from each other to avoid premature convergence to a suboptimal solution. DROO designs the quantizer following an order-preserving rule to generate high quantization diversity \cite{Huang2020}. We can further introduce random noise into the quantizer to increase exploration \cite{Yan2020a}.\
	
	\item \textbf{Optimization-based Critic Module}: Instead of using another DNN in the critic module as in conventional actor-critic frameworks, DROO takes advantage of the model information to precisely assess the offloading decisions through model-based optimization. In particular, the critic module evaluates the $K$ binary decisions $\bm x_j^t$'s by solving the corresponding resource allocation subproblem and obtains the score $\tilde{f}\left(\bm s^t, \bm x_j^t\right) = f\left(\bm s^t, \left\{\bm x_j^t, \bm y^{t\ast}\left(\bm x_j^t\right)\right\}\right)$ for each $\bm x_j^t$, where $\bm y^{t\ast}\left(\bm x_j^t\right)$ is the optimal resource allocation given $\bm x_j^t$. Then, the action selection step selects the offloading decision with the highest score, and outputs the action $\bm{a}^t=\left\{\bm x^t,\bm y^{t\ast}\left(\bm x^t\right)\right\}$. The action is then executed in the current time slot. Notice that in most cases, the resource allocation subproblem is convex, so that we can obtain the optimal $\bm y^{t\ast}\left(\bm x^t\right)$ efficiently. From the above description, the major computation complexity of DROO arises from solving the resource allocation subproblem $K$ times to acquire the best $\bm x^t$ in each time slot. {\color{black}{A larger $K$ explores more binary decisions, which leads to faster convergence rate and better offloading policy of the DNN after convergence. However, a larger $K$ also incurs heavier computation complexity in the online offloading decision process. To balance the performance and computation complexity, DROO initially sets a large $K$ and adaptively decreases $K$ over time as the DNN gradually approaches the optimal policy \cite{Huang2020}.}} By doing so, DROO incurs very low computational complexity even with a relatively large action space.
	
	\item \textbf{Policy Update Module}: DROO follows the standard experience replay technique to update the policy of the actor module. Specifically, the best decision $\bm x^t$ and the corresponding input $\bm s^t$ are added as a newly labeled training sample to a replay memory, from which a batch of samples is randomly selected to update the parameters of the DNN at every training interval. When the replay memory is full, the oldest training samples will be overwritten by the newest ones.
\end{itemize}

As iterations proceed, DROO gradually approaches the optimal state-action mapping by learning from the best actions experienced most recently. After convergence, DROO can quickly output the optimal solution instead of computing the COP from scratch when observing new input problem parameters. \

DROO significantly improves the convergence of the training process compared with conventional actor-critic DRL, thanks to the following two reasons. First, the COP is decoupled under the learning-optimization structure. Both the model-free actor module and the model-based critic module solve much simpler sub-problems than the original large-size MINLP. In particular, the actor module solves a conventional multi-class classification problem, and the critic module tackles an ``easy'' continuous optimization. Second, model-based optimization in the critic module not only guarantees the feasibility of the output action, but also provides quick, precise, and stable evaluation of the actions. This fundamentally improves the convergence of the DRL algorithm.\

To evaluate the performance of DROO, we conduct numerical simulations in a multi-user MEC network. We define the reward of each MD as the computation rate (i.e., the total amount of task data processed at both MDs and ES) in a time slot and the system utility as the weighted sum computation rate of all MDs. {\color{black}{We aim to maximize the system utility by treating the wireless channel gains as the input problem parameters. In the simulations, we set the weighting factor of the $i$th MD as 1 if $i$ is odd and as 1.5 otherwise.}} We set the actor DNN as a fully-connected multi-layer perceptron consisting of one input layer, two hidden layers, and one output layer, where the first and the second hidden layers have $120$ and $80$ hidden neurons, respectively. The two hidden layers use ReLu activation function, and the output layer uses sigmoid activation function. The replay memory size $J = 1024$ and the training sample batch size is $128$.{\footnote{The complete source code implementing DROO is available on-line at https://github.com/revenol/DROO.}}\

\begin{figure}
	\centering
	\includegraphics[scale=0.53]{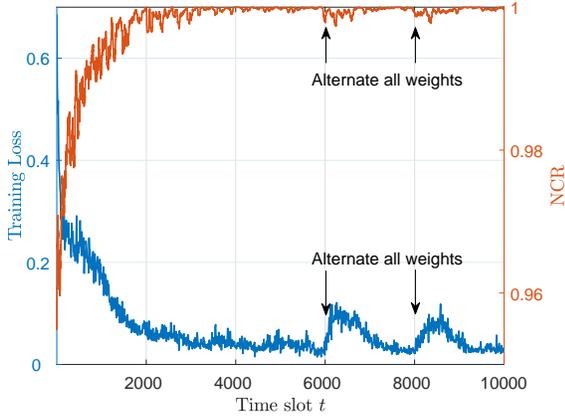}
	\captionsetup{font=footnotesize}
	\caption{Convergence performance of DROO.}
	\label{fig4} 
\end{figure}

\begin{figure}
	\centering
	\includegraphics[scale=0.55]{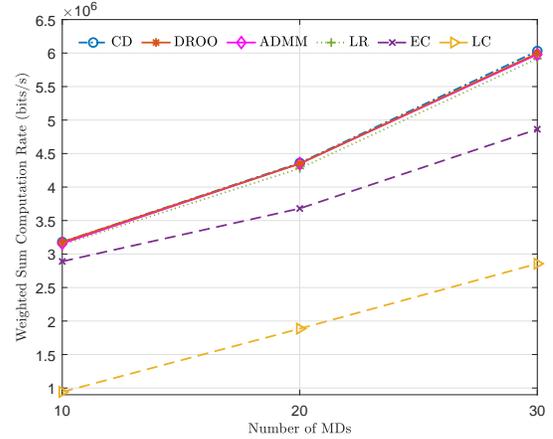}
	\captionsetup{font=footnotesize}
	\caption{Computation rate performance comparison.}
	\label{fig42} 
\end{figure}

In Fig. \ref{fig4}, we first investigate the training loss and normalized computation rate (NCR) of DROO with $10$ MDs, where the training loss is computed by the averaged cross-entropy loss function and the computation rate is normalized by the maximum achievable rate. As shown in the figure, DROO only requires about $t=3000$ slots to converge. The computation rate approaches the optimal one at convergence. To further verify the robustness of DROO, we alternate the values of all weighting factors (from 1 to 1.5 and vice versa) at both $t=6000$ and $t=8000$. The results show that DROO quickly adapts to the new parameters and automatically converges to new optimum even with a sharp variation of the weighting factors.

{\color{black}{In Fig. \ref{fig42}, we compare the computation rate of DROO with five representative benchmark algorithms when the number of MDs varies. Specifically, local computing (LC) refers to the case where all MDs compute locally. Likewise, edge computing (EC) enforces the MDs to offload all the tasks for edge computing. Meanwhile, we plot the result of CD (coordinate descent) as an optimal benchmark, as CD is likely to achieve a close-to-optimal solution \cite{Bi2018a}. The results show that DROO achieves almost identical performance as the CD method, and significantly outperforms the LC and EC methods. To show the real-time performance of DROO, we compare in \cite{Huang2020} the computational time for LR (linear relaxation), ADMM, CD and DROO to output a solution of $\bm{a}^t$. The training of the DNN is infrequent (once every tens of time slots) and can be performed in parallel with task offloading and computation, which does not incur additional delay overhead. Therefore, we only consider the computation delay of offloading action generation when comparing the computation delay with other benchmark methods. The results in \cite{Huang2020} show that DROO requires substantially less computational time than LR, ADMM, and CD (only about 0.059s with 30 MDs compared to 0.81s for LR, 3.9s for ADMM and 3.8s for CD methods). This makes DROO technically viable for real-time offloading control in a fast-varying edge environment.}}

\section{A Lyapunov-aided DRL Framework For the Stochastic COP}\label{sec4}
\begin{figure}
	\centering
	\includegraphics[scale=0.5]{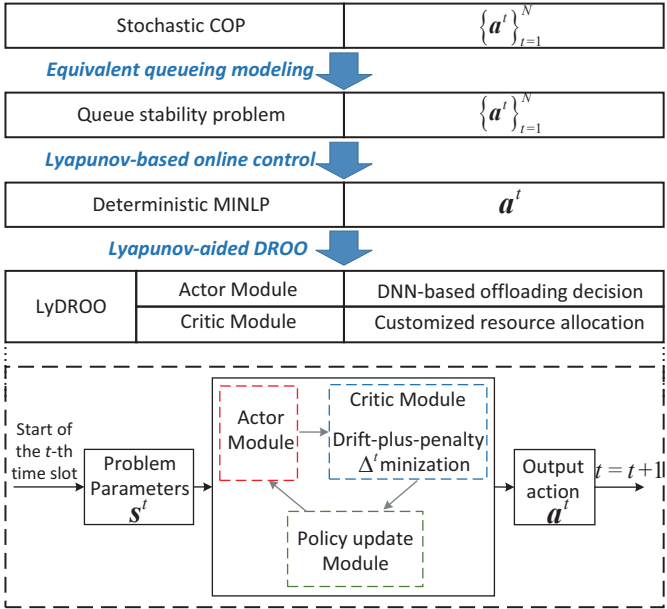}
	\captionsetup{font=footnotesize}
	\caption{Illustration of LyDROO for solving multi-slot COP.}
	\label{fig5}
\end{figure}

In this section, we extend the basic DROO framework to solve a stochastic COP with both long-term objective and constraints. The key is to integrate the well-known Lyapunov optimization in the design of the critic module, resulting in a Lyapunov-aided DROO framework named LyDROO \cite{Bi2021}. As shown in Fig. \ref{fig5}, LyDROO makes three major modifications to the basic DROO framework.\

\begin{itemize}[leftmargin=*]
	\item{\textbf{Equivalent Queueing Modeling}}: Besides considering the physical queue backlogs, such as task data and battery energy queues, LyDROO introduces virtual queues for the long-term constraints such as average energy consumption and computation delay. {\color{black}{Based on the Lyapunov optimization theory, LyDROO interprets the stochastic COP as an equivalent queue stability control problem, where the long-term constraints can be satisfied by stabilizing all the actual and virtual queues $\bm{I}^t$.}}
	
	\item{\textbf{Lyapunov-based Online Control}}: LyDROO adopts Lyapunov optimization to remove the temporal correlations among actions in different time slots while keeping stable $\bm I^t$. {\color{black}{Specifically, LyDROO slightly modifies the critic module of DROO in Fig. \ref{fig3}. That is, instead of maximizing $\tilde{f}$ in the critic, it minimizes a drift-plus-penalty function $\Delta^t$ for each candidate binary candidate action, where $\Delta^t$ increases with $\bm I^t$ and decreases with the utility value $f$ in the $t$th time frame.}} Intuitively, by constantly minimizing $\Delta^t$, LyDROO simultaneously maximizes the long-term utility and satisfies the long-term constraints by keeping all the queues stable. {\color{black}{Notice that minimizing $\Delta^t$ requires only the problem parameters of the $t$th time frame. Therefore, we can implement the proposed LyDROO algorithm in a fully online manner without any future information.}}.		
	
	\item{\textbf{Lyapunov-aided Critic}}: As illustrated at the right half of Fig. \ref{fig5}, LyDROO solves each per-slot deterministic MINLP following the idea of DROO. At the begining of the $t$th time slot, LyDROO inputs the current problem parameters $\bm{s}^t$ to the actor module, which outputs $K$ candidate binary offloading decisions. The critic module optimizes the resource allocation variables by minimizing the drift-plus-penalty function $\Delta^t$ given the offloading decisions. After obtaining the scores of all the $K$ offloading actions, LyDROO stores the best-scored offloading action $\bm{x}^t$ and the input $\bm{s}^t$ as a new training sample into the replay memory to update the policy of the actor module. Then, LyDROO renews the queue state $\bm{I}^t$ by executing the offloading action $\bm{x}^t$ in the current time slot and starts a new iteration until convergence.	
\end{itemize}

In the following, we conduct numerical simulations to evaluate the performance of LyDROO. We consider an MEC network of $10$ MDs with random task data arrivals. We intend to maximize the time-averaged weighted sum computation rate of all MDs while guaranteeing the task data queue stability and average power constraints at each MD. {\color{black}{The input problem parameters include the wireless channel gains from all MDs to the ES, the task data arrivals at all MDs, and the data queue and virtual energy queue backlogs at all MDs.}} {\color{black} We assume that the task data arrival rates of all MDs follow exponential distribution with identical average rate $\lambda$. We set the average power constraint at each MD to 0.08 Watt.} {\color{black}{We consider two different deep learning models in the actor module of LyDROO: the DNN network and the CNN network.{\footnote{The complete source code implementing LyDROO is available on-line at https://github.com/revenol/LyDROO.}} For performance comparison, we also consider two benchmark methods: Lyapunov-guided coordinated decent (LyCD) and Myopic optimization.}} Similar to LyDROO, LyCD transforms the stochastic problem into the per-slot deterministic MINLP, but solves the deterministic problem using optimization-based CD method instead of DROO. LyCD yields close-to-optimal performance, and yet is unsuitable for real-time implementation due to its high computational complexity. On the other hand, the Myopic method ignores the queue backlogs and greedily maximizes the weighted sum computation rate in each time slot until the prescribed energy budget exhausts.\

\begin{figure}
	\centering
	\includegraphics[scale=0.55]{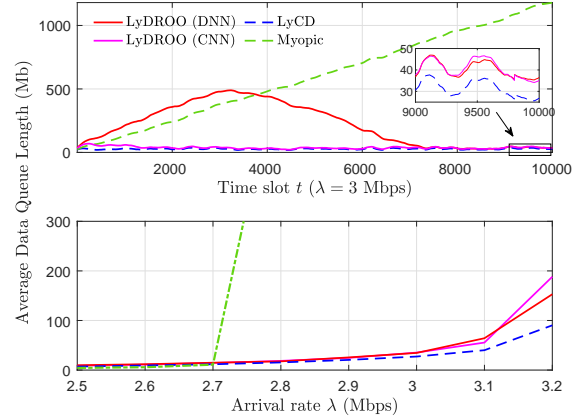}
	\captionsetup{font=footnotesize}
	\caption{Average data queue length under different task arrival rate $\lambda$. }
	\label{fig6} 
\end{figure}

The simulation results are plotted in Fig. \ref{fig6}. In the top sub-figure, we plot the variation of average data queue length of the $10$ MDs over time with $\lambda=3$ Mbps. The results show that the Myopic method is unable to stabilize all the data queues. {\color{black}{In contrast, the queue length of LyDROO becomes stable as time proceeds. Compared to the DNN network, the CNN network converges much faster and significantly reduces the data queue backlog during the learning stage.}} In the bottom sub-figure, we compare the average queue length of the three methods under different $\lambda$. The results show that, LyCD and LyDROO yield stable data queues for all considered data arrival rates. In contrast, Myopic cannot maintain a stable data queue when $\lambda > 2.7$ Mbps. {\color{black}{It is also worth mentioning that both DNN- and CNN-based LyDROO have similar stable computation rate regions as LyCD, but taking much less computational time (e.g., 0.156s for DNN-based LyDROO vs. 8.02s for LyCD with 30 MDs.) \cite{Bi2021}.}} These numerical results verify the effectiveness of LyDROO on solving the online offloading problem with long-term constraints.


\section{Future research directions}\label{sec5}
In this section, we outline several future research directions that we find particularly valuable.
\vspace{0 mm}
\subsection{Online Offloading Control under Future Prediction}
In a stochastic MEC system, the decision maker can predict future system information (e.g., locations of moving MDs) based on historical observation to assist online offloading decisions. One possible solution is to modify the actor module of DROO, e.g., by replacing the DNN with other neural networks with prediction capability such as recursive neural network (RNN) and long short term memory (LSTM) network. Besides, we may incorporate robust optimization techniques in the critic module to alleviate the negative impact of imprecise prediction.

\subsection{Distributed Online Offloading Control}
When implementing DROO and LyDROO, the system operator makes centralized control based on the global system information collected from all MDs. Centralized control, however, could be costly in a large distributed MEC system. Meanwhile, due to the security and privacy concerns, MDs may be reluctant to share the parameters containing private data (e.g., device parameters which may expose user's shopping habits). To address this problem, a promising approach is to exploit distributed learning/optimization techniques such as federated learning and multi-agent learning to design individual actor and critic modules at distributed MDs.

\subsection{Applications beyond Online Computation Offloading}
Beyond solving the COPs in MEC networks, the DROO framework is applicable to online control in a wide-range of applications with hybrid integer-continuous optimization structure. As shown in Fig. \ref{fig1}, the binary decisions may denote the routing planning in an IoT network, mode selection in a device-to-device assisted communication system, user association in a UAV-assisted scenario, or service placement in a vehicular environment. Besides, DROO can be applied to on-off beamforming control in Intelligent reflecting surface (IRS) and massive multiple-input and multiple-output (MIMO) systems, etc.

\section{Conclusion Remarks}
In this article, we have provided an overview of the online computation offloading control methods in MEC systems. In particular, we introduced an integrated optimization-learning framework, DROO, that takes advantage of both past experience and model information to provide fast and robust convergence as well as close-to-optimal real-time offloading control. By incorporating the Lyapunov optimization, we demonstrated the flexibility of DROO to solve long-term stochastic online control problems. Besides, we have highlighted several valuable future research topics and discussed the challenges therein.

\ifCLASSOPTIONcaptionsoff
  \newpage
\fi
\bibliographystyle{IEEEtran}
\bibliography{IEEEabrv,MyRefLib}
\vspace{3 mm}
\section*{Biographies}
\vspace{-12 mm}
\begin{IEEEbiographynophoto}{Xian Li [M'20]}
is now a postdoctoral research fellow with the College of Electronics and Information Engineering, Shenzhen University, China. His research interests mainly include optimizations in mobile edge computing and wireless powered communication networks.
\end{IEEEbiographynophoto}
\vspace{-10 mm}
\begin{IEEEbiographynophoto}{Liang Huang [M'16]}
	is now an Associate Professor with the College of Computer Science and Technology, Zhejiang University of Technology, China. His research interests lie in the areas of queueing and scheduling in communication systems and networks.
\end{IEEEbiographynophoto}
\vspace{-10 mm}
\begin{IEEEbiographynophoto}{Hui Wang}
	is now a Professor with the Shenzhen Institute of Information Technology. His research interests include wireless communication, signal processing, and distributed computing systems.
\end{IEEEbiographynophoto}
\vspace{-10 mm}
\begin{IEEEbiographynophoto}{Suzhi Bi [M'14, SM'19]}
	is now an Associate Professor with the College of Electronics and Information Engineering, Shenzhen University, China, His research interests include the optimizations in wireless information and power transfer, mobile computing, and smart power grid communications.
\end{IEEEbiographynophoto}
\vspace{-10 mm}
\begin{IEEEbiographynophoto}{Ying-Jun Angela Zhang [M'05, SM'11, F'19]}
	is now a Professor with the Department of Information Engineering, The Chinese University of Hong Kong. Her research interests focus on optimization and learning in wireless communication systems. She is a Fellow of IEEE, a Fellow of IET, and an IEEE ComSoc Distinguished Lecturer.
\end{IEEEbiographynophoto}






\end{document}